# On melting of boron subnitride $B_{13}N_2$ under pressure

Vladimir L. Solozhenko* and Vladimir A. Mukhanov

*LSPM–CNRS, Université Paris Nord, 93430 Villetaneuse, France*

E-mail: vladimir.solozhenko@univ-paris13.fr

Melting of rhombohedral boron subnitride $B_{13}N_2$ has been studied *in situ* at pressures to 8 GPa using synchrotron X-ray diffraction and electrical resistivity measurements. It has been found that above 2.6 GPa $B_{13}N_2$ melts incongruently, and the melting curve exhibits positive slope of 31(3) K/GPa that points to a lower density of the melt as compared to the solid phase.

*Keywords:* boron subnitride, melting, high pressure, high temperature.

Rhombohedral boron subnitride $B_{13}N_2$ has been recently synthesized at 5 GPa by crystallization from melt of the B-BN system [1-3]. The structure of $B_{13}N_2$ belongs to the *R-3m* space group and represents a new structural type produced by the distorted $B_{12}$ icosahedra linked by N–B–N chains and inter-icosahedral B–B bonds [1]. $B_{13}N_2$ is low-compressible ($B_0$ = 200(15) GPa [4]) boron-rich solid and is expected to exhibit hardness of 40.3 GPa [5,6] i.e. belongs to the family of superhard phases. The thermal stability and melting behavior of $B_{13}N_2$ was not studied so far, even at ambient pressure. Here we present the first experimental data on melting of boron subnitride at pressures to 8 GPa.

Polycrystalline $B_{13}N_2$ has been synthesized in a toroid-type apparatus at 5 GPa by quenching of the B–BN melt from 2630 K in accordance with high-pressure phase diagram of the B–BN system [7]. Powders of crystalline β-rhombohedral boron (99%, Alfa Aesar) and hexagonal graphite-like boron nitride (hBN) (99.8%, Johnson Matthey GmbH) have been used as starting materials. The X-ray diffraction study (TEXT-3000 Inel, CuKα1 radiation) has shown that the quenched samples contain well-crystallized $B_{13}N_2$ (*a* = 5.4585(8) Å, *c* = 12.253(2) Å), usually with traces of β-B and hBN due to the peritectic nature of the L + BN ⇌ $B_{13}N_2$ reaction [7]. To remove boron impurity, the samples were ground and treated with 7N nitric acid (ACS, Alfa Aesar) for 20 min at 370 K, washed with deionized water and dried at 400 K.

$B_{13}N_2$ melting in the 4.3–5.6 GPa pressure range was studied by energy-dispersive synchrotron X-ray diffraction using MAX80 multianvil system at F2.1 beamline, DORIS III storage ring (DESY). The experimental details are described elsewhere [8,9]; the results are shown in Figure.

Melting of boron subnitride in the 2.6–7.7 GPa range was studied *in situ* by electrical resistivity measurements [10] in a specially designed high-temperature cell [11] of a toroid-type high-pressure apparatus. The experimental details are described elsewhere [12]. No signs of chemical interaction between $B_{13}N_2$ and graphite electrical inputs were observed. The experimental data are presented in Figure.

Melting of boron subnitride is accompanied by a drop in electrical resistance that is reproducible upon several heating/cooling cycles. The specific electrical conductivity of the melt at 6.1 GPa and 2640 K was evaluated as $2.4(9) \times 10^5\ \Omega^{-1} \cdot m^{-1}$, that is very close to the conductivity of boron melt at the same *p-T* conditions [10], while conductivity of solid $B_{13}N_2$ at this pressure ($2.3(11) \times 10^4\ \Omega^{-1} \cdot m^{-1}$ at 2590 K) is lower by an order of magnitude.

The $B_{13}N_2$ melting curve (dashed line obtained by the least-squares method from the results of electrical resistivity experiments) exhibits positive slope 31(3) K/GPa, which points to a lower density of the B–N liquid as compared to solid $B_{13}N_2$. Extrapolation of the melting line to the low-pressure region allows estimating the melting point of $B_{13}N_2$ at ambient pressure as 2430(20) K that is slightly higher than the melting temperature of β-rhombohedral boron (2360 K [13]).

In all experiments crystallization of the B–N liquid resulted in formation of $B_{13}N_2$ in mixture with boron nitride (hBN or cBN depending on pressure) and novel tetragonal boron subnitride $B_{50}N_2$ [3] (space group *P-4n2*; $a = 8.8044(7)$ Å, $c = 5.0395(6)$ Å) that is indicative of the incongruent type of $B_{13}N_2$ melting in the pressure range under study. According to *in situ* X-ray diffraction data, at pressures to 8 GPa the icosahedral crystal structure of $B_{13}N_2$ remains stable up to the melting similar to other icosahedral boron-rich solids, namely, $B_6O$ [14] and $B_4C$ [11].

**Acknowledgments.** We thank Prof. V.Z. Turkevich for valuable discussions. Synchrotron X-ray diffraction experiments were carried out during beam time allocated to Project DESY-D-I-20090007 EC at DESY. This work was financially supported by European Union's Horizon 2020 Research and Innovation Programme under Flintstone2020 project (grant agreement No 689279).

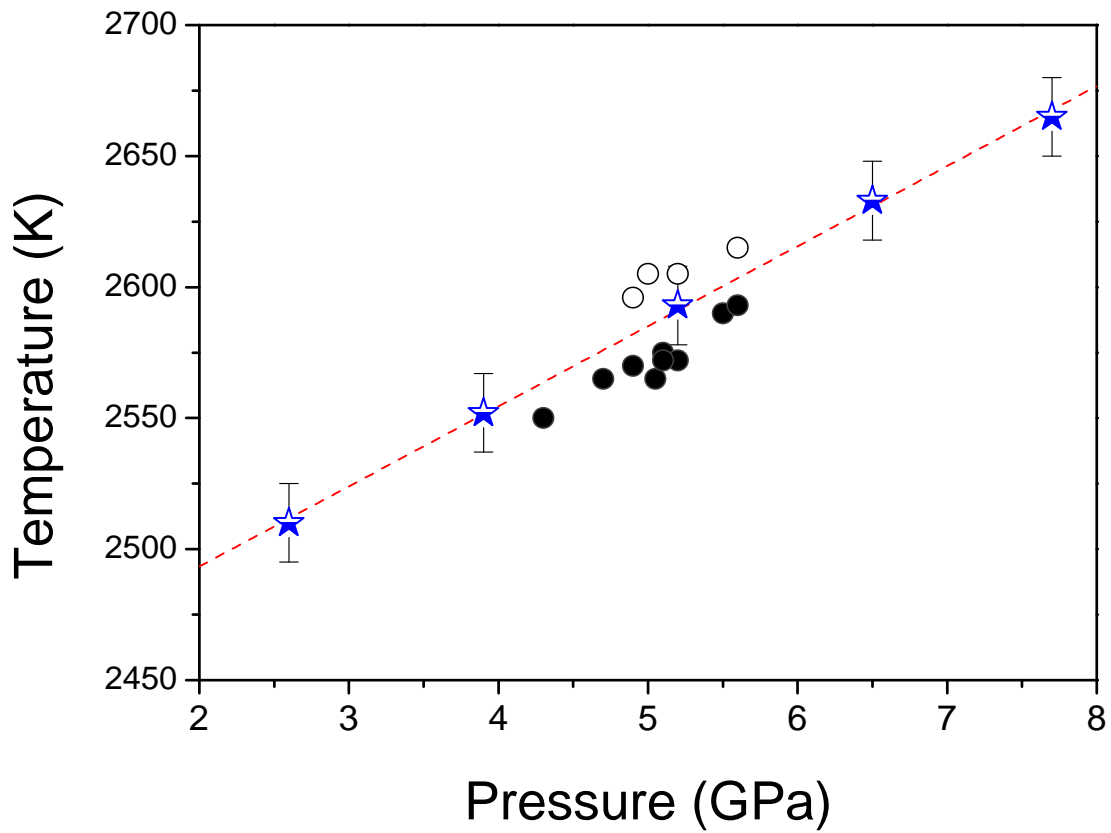

Pressure dependence of $B_{13}N_2$ melting temperature. The results of synchrotron X-ray diffraction experiments are presented by circles (solid symbols correspond to solid $B_{13}N_2$, open symbols – to the melt). Half-filled stars show the onset of melting registered *in situ* by electrical resistivity measurements; dashed line is the linear approximation of the experimental data.